\documentclass[twocolumn,showpacs,preprintnumbers,amsmath,amssymb]{revtex4}
\usepackage{graphicx}% Include figure files
\usepackage{dcolumn}% Align table columns on decimal point
\usepackage{bm}% bold math
\begin{document}
%\draft

\title{Continuous Charge Modulated Diagonal Phase in Manganites.}

\author{Luis Brey }

\affiliation{\centerline {Instituto de Ciencia de Materiales de
Madrid (CSIC),~Cantoblanco,~28049~Madrid,~Spain.}} \
%\maketitle
\begin{abstract}

We present a novel ground state that explain the continuous
modulated charge diagonal order recently observed in manganese
oxides, at hole densities $x$ larger than one half. In this
diagonal phase the charge is modulated with a predominant Fourier
component inversely proportional to $1-x$. Magnetically this state
consist of antiferromagnetic coupled zig-zag chains. For a wide
range of relevant physical parameters as electron-phonon coupling,
antiferromagnetic interaction  between Mn ions and on-site Coulomb
repulsion, the diagonal phase is the ground state of the system.
The diagonal phase is favored by the modulation of the hopping
amplitude along the zig-zag chains, and it is stabilized with
respect to the one dimensional straight chain by the electron
phonon coupling. For realistic estimation of the physical
parameters, the diagonal modulation of the electron density is
only a small fraction of the average charge, a modulation much
smaller than the obtained by distributing Mn$^{+3}$ and Mn$^{+4}$
ions. We discuss also the spin and  orbital structure properties
of this new diagonal phase.

\end{abstract}


\maketitle

Oxides of type (R$_{1-x}$A$ _x$)MnO$_3$ where R denotes rare earth
ions and A is a divalent alkaline ion, are called generically
manganites. In these compounds $x$ coincides with the
concentration of holes moving in the $e_g$ orbital band of the Mn
ions that ideally form a cubic structure  of lattice parameter
$a$. The ground state (GS) properties of manganites are determined
by the competition between at least four independent energy
scales; the anti-ferromagnetic (AF) interaction between  the  Mn
spins, the electron phonon coupling, the electronic repulsion and
the kinetic energy of the carriers. In manganites the energy
magnitude of these effects is the same and very different states
can have very similar energies. Consequentially by varying
slightly parameters as carrier concentration, strain, disorder,
temperature etc, different GS as ferromagnetic metallic
phases\cite{Wollan}, AF Mott insulator\cite{Kanamori}, charge and
orbital ordered stripe phases\cite{Chen1,Chen2,Mori,Chen3}, or
ferromagnetic charge ordered phases\cite{Loudon1} can be
experimentally observed.

Studies  based in pure double-exchange (DE)
model\cite{Calderon1,Alonso}, electron-phonon
coupling\cite{Dagottobook}, orbital ordering\cite{Brink},  strain
interactions\cite{Calderon2}, or \textit{ab initio}
calculations\cite{Ferrari} have been successfully applied for
explaining the existence of ferromagnetic metallic phases at $x
\sim$ 0.2-0.4 and the existence of the CE phase at half doping.
Also for large enough AF coupling between the Mn spins orbital
ordering could induce stripes at $x$=$1/m$ being $m$ an
integer\cite{Hotta1,Hotta2}. In the CE phase the $x$-$y$ layers
are AF coupled and into the plains the magnetic structure is
formed by AF coupled  zig-zag chains. The horizontal ($x$) and
vertical ($y$) steps of the zig-zag chains contain three Mn ions.
In the CE phase the charge is stacked in the $z$-direction whereas
in the $x$-$y$ planes  it is ordered in a checkerboard form.

For $x>$1/2 microscopic models\cite{Brink,Maitra} predict the
existence of a C phase formed by vertical one dimensional chains
coupled AF. However experiments reveal the existence of a diagonal
charge modulation, that was interpreted\cite{Mori,Chen3}  as a
mixture of two commensurate adjacent integer period sub-units. On
the other hand, by similitude with ferroelectric it has been argue
that for $x>$1/2 the GS could be formed by commensurate regions of
density $x$=$(1- \frac{1}{n})$ separated by solitons where the
extra charge is localised\cite{Mathur}. In this picture the
stripes are formed by an array of solitons. However recently, for
doping $x>$0.5, low temperature electron microscopy
experiments\cite{Loudon2} shown an uniform periodicity
proportional to $1-x$, ruling out the existence of a mixture of
integer  period units joined by discommensurations or solitons.
Also at $x$=2/3 a diagonal supercell structure has been proposed
for modelling neutron scattering data\cite{Fernandez}.

In this work we propose a new GS for the manganites at doping
$x$=$(1-\frac{n}{m})>$0.5. In this phase the charge is stacked in
layers and at each plane it is modulated diagonally with only a
predominant Fourier component $(\frac{2\pi}{a}
,\frac{2\pi}{a})\frac{m}{n}$. Magnetically this \textit{diagonal}
phase consists of AF coupled $x$-$y$ layers  formed of zigzag
chains also coupled AF. The horizontal and vertical steps of the
chain are formed by $(m+1)$ Mn ions. By varying the values of $n$
and $m$ this phase describe a GS where the periodicity of the
charge changes continuously  with $x$.

\textit{Model.} In ideal manganites the Mn ions form a cubic
lattice. The crystal field split the Mn $d$ levels into an
occupied  strongly localised $t_{2g}$ triplet and a doublet of
$e_g$ symmetry. Coulomb repulsion prevents double occupancy and
aligns the spins of the $d$ orbitals. At  $x$=1 the $e_g$ orbitals
are empty and the superexchange coupling  between the Mn  spins
produces an AF GS. At $x \neq $1 the hopping of electrons between
empty $e_g$ states is possible. The Hund's coupling between the
spins of the carriers and each core spin is much larger than the
kinetic energy, and each electron is forced to align with the core
spin texture. Then the carriers can be treated as spinless
particles and the  hopping amplitude between two Mn ions is
modulated by the factor $f_{1,2}= \cos\frac{\vartheta
_1}{2}\cos\frac{\vartheta _2}{2} + e ^{i ( \phi _1 - \phi _2)}
\sin\frac{\vartheta _1}{2}\sin\frac{\vartheta _2}{2}$ where
$\{\vartheta _i, \phi _i \}$ are the Euler angles of the, assumed
classical,  Mn core spins $\{\textbf{S} _i \}$ . This is the so
called  DE model

For obtaining the GS of the system we study a three dimensional DE
model coupled to Jahn-Teller (JT) phonons. We also include the AF
coupling between the Mn core spins $J_{AF}$ and a Hubbard term $U$
for describing the strong inter-orbital Coulomb interaction.
\begin{eqnarray}\label{Hamiltonian}
 H &  = &  -\sum_{i,j,a,a'} f _{i,j} t _{a,a'} ^{u} C ^+ _{i,a} C
_{j,a'} +U \sum _{i,a,a'} n _{i,a} n _{i,a'}
\nonumber \\
 &+& J_{AF} \sum _{<i,j>}
\textbf{S} _i \textbf{S} _j
 + \frac{1}{2} \sum _i \left ( \beta Q _{1i} ^2+ Q
^2_{2i} + Q _{3i} ^2 \right )
\nonumber \\
 & + & \lambda \sum _{i} \left ( Q _{1i} \rho _i + Q _{2i} \tau _{xi}
+ Q _{3i} \tau _{zi} \right ) \, \, \, ,
\end{eqnarray}
here $C^+ _{i,a}$ creates an electron in  the Mn ions located at
site $i$ in the $e_g$ orbital $a$ ($a$=1,2 with 1=$|x^2-y^2>$ and
2=$|3z^2-r^2>$). The hopping amplitude is finite for next
neighbors  Mn and depends both on the type of orbital involved and
on the direction $u$ between sites $i$ and $j$
($t_{1,1}^{x(y)}=\pm \sqrt{3} t_{1,2}^{x(y)} =\pm \sqrt{3}
t_{2,1}^{x(y)}=3t_{2,2}^{x(y)}=t$ and $t_{2,2}^{z}=4/3 t$ with
$t_{2,2}^{z}=t_{2,1}^{z}=t_{1,2}^{z}=0$)\cite{Dagottobook}.In the
rest of the paper $t$ is taken as the energy unit. In the second
term of Eq.(\ref{Hamiltonian}) $n_{i,a} = C^{+}_{i,a} C _{i,a}$.
The fifth term couples the $e_g$ electrons with the three active
MnO$_6$ octahedra distortions: the breathing mode $Q_{1i}$, and
the JT modes $Q_{2i}$ and $Q_{3i}$ that have symmetry $x^2$-$y^2$
and $3z^2$-$r^2$ respectively. $Q_{1i}$ couples with the charge at
site $i$,  $\rho _i = \sum _a C^+ _{i,a} C _{i,a}$ whereas
$Q_{2i}$ and $Q_{3i}$ couple with the $x$ and $z$ orbital
pseudospin density , $\tau _{xi}= C^+ _{i1}C_{i2} + C^+ _{i2}
C_{i1}$ and $\tau _{zi}= C^+ _{i1}C_{i1} - C^+ _{i2} C_{i2}$,
respectively. $\lambda$ is the electron-phonon coupling constant.
The forth term is the elastic energy of the octahedra distortions,
being $\beta \geq$2 the spring constant ratio for breathing and
JT-modes\cite{Aliaga}. In the perovskite structures the oxygens
are shared by neighbors MnO$_6$ octahedra and the $Q$'s
distortions are not independent, being cooperative effects very
important\cite{JAV}. In order to consider this collective effect
we consider the position of the oxygen atoms as the independent
variables of the JT distortions.

For a particular electron density,  a given set of values {$
\lambda$, $J_{AF}$ and $U$}, and a texture of core spins $\{
\textbf{S}_i \}$, we solve self-consistently the mean field
version of Hamiltonian (\ref{Hamiltonian}) and obtain the energy,
the local charges $\{\rho_i\}$ and orbital pseudospin orientation
$\{\tau _{xi}, \tau_{zi} \}$. We have solved the Hamiltonian for
$x\leq 0.5$ and with the appropriated parameters  we have
recovered the GS obtained previously by different
groups\cite{Hotta1,Hotta2,Aliaga,Brink,Alonso,Efremov}. Therefore
we are in conditions for studying the different phases that can
appear for $x>0.5$.

\textit{Results.} We have studied different magnetic texture
states, as the three dimensional (3D) ferromagnetic (FM) phase,
the A phase formed by FM two dimensional (2D)  planes coupled AF,
the C and the CE phases, the island phases\cite{Aliaga1,Alonso} of
the form $(\pi/3,\pi,\pi)$, the $C_xE_{1-x}$ phases\cite{Hotta2},
the Skyrmion phases\cite{Alonso} and the 3D AF phases. Neither of
the GS's obtained by comparing the energy of these magnetic phases
present a diagonal modulation of the charge, in any range of the
parameters $\{\lambda, J_{AF}, U\}$. Neither of them present a
continuous variation of the period of the charge modulation when
the hole concentration varies.

In this work  we propose, for $x$=$(1-\frac{n}{m})$, a new
magnetic phase consisting of AF ordered layers composed of AF
coupled zig-zag chains, with vertical and horizontal steps formed
by $m+1$ Mn ions. As the tunnelling amplitude depends on the
direction between sites ($t_{1,2}^x$=$- t_{1,2}^y$) in this phase
the hopping amplitude is modulated with period 2$m$  along the
zig-zag chains, and this make possible the existence of Fourier
components of the charge of the form $(\frac{2\pi}{a}
,\frac{2\pi}{a})\frac{m}{n}$. Even in absence of interactions, the
modulation of the hopping along the chain   produces the
appearance of gaps in the energy spectra at hole concentrations
$1-\frac{n}{m}$ and the system behaves as a band
insulator\cite{nota3}. In the case of $m$=2 this phase coincides
with the CE phase When the Hubbard term or the electron-phonon
coupling is included, the gap at the hole density $1-\frac{n}{m}$
is enlarged and the charge density modulation with period
$\frac{m}{n}(a,a)$ is privileged, being the charge modulated
diagonally  basically  with a single Fourier component.

\begin{figure}
  \includegraphics[clip,width=8cm]{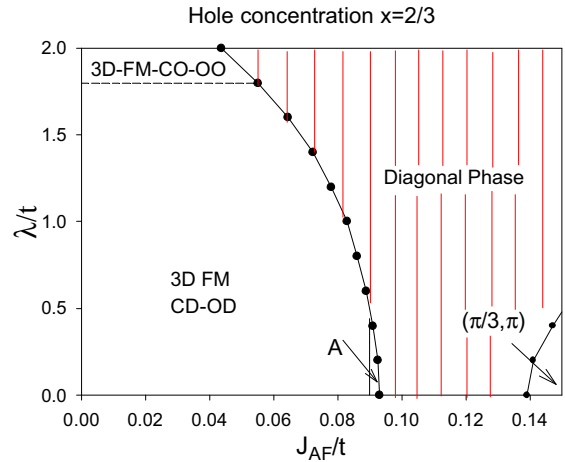}
  \caption{(Color online) Phase diagram for $x$=2/3 for the three-dimensional DE
  two orbital model with cooperative Jahn Teller phonons. The on-site Coulomb term $U$ is
  take as zero. The symbols CO(CD)
   and OO(OD) stand for charge and orbital ordered(disordered) respectively.}
\label{fig1}
\end{figure}

The unit cell of this phase contains 4$m \times$4$m$ Mn ions and
this impose a computational restriction to the studied hole
densities. We have calculated, for $x$=2/3 and $x$=3/5, the energy
of the diagonal and other established phases for  different values
of $\lambda$, $U$ and $J_{AF}$. By comparing energies we build the
corresponding phase diagrams. In Fig.1 we plot the $\lambda -
J_{AF}$ phase diagram for $x=2/3$ and $U=0$. At small values of
$J_{AF}$ and $\lambda$ the GS is the uniform charge 3D FM phase,
this state is the standard FM metallic phase that appear in the
canonical DE model. At small values of $J_{AF}$ and large values
of $\lambda$ the GS is an insulating 3D FM phase where the
electric charge and the orbital are ordered. This charge ordered
FM phase is the 3D analog to the 2D charge ordered phase reported
in reference \cite{Yunoki}. At moderate values of $J_{AF}$ and in
a wide range of values of $\lambda$ there is a large window in the
phase diagram where the GS  is the diagonal phase. At small
$\lambda$ the 3D-FM state is separated from the diagonal phase by
the A phase. For large values of $J_{AF}$ island phases of the
form $(\pi /3 ,\pi)$  appear\cite{Aliaga} and for much larger
values of $J_{AF}$ (not shown in Fig.1) the pure AF phase wins in
energy. Topologically equivalent phase diagram is obtained for
$x=3/5$. The important point to note  is that the window of
$\lambda$-$J_{AF}$ values, where the diagonal phase is stable
includes the values for which the CE phase is the GS at $x$=1/2.
Therefore we expect  the diagonal phase to be the GS for $x>$1/2.

\begin{figure}
  \includegraphics[clip,width=8cm]{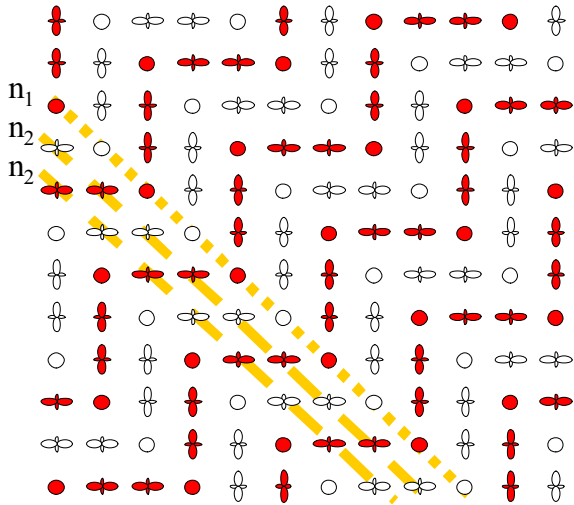}
  \caption{(Color online) Schematic view of the diagonal phase in the $x$-$y$ plane.
  Open and solid symbols indicates up and down $t_{2g}$ spins respectively.}
\label{fig2}
\end{figure}

In Fig.2 we plot schematically the spin-orbital ordering in the
unit cell  of the diagonal phase at $x$=2/3. Open and solid
symbols represent up and down Mn core spins. The lobes indicates
that the electron occupy $3x^2-r^2$ or $3y^2-r^2$ orbitals whereas
small circles represent the occupation of the $x^2-y^2$ orbital.
The charge density is constant along diagonal directions marked in
Fig.2 and periodic along the perpendicular direction. At x=2/3 the
electric charge on the Mn ions have only two  values $n_1$ or
$n_2$, and the charge disproportionation  of the electron charge
depends on $\lambda$ and $U$. For realist values of
$\lambda$=0.6$t$ and $U$=4$t$ the difference of charge is
$n_1$-$n_2$=0.15. This value is very different from the extreme
value of 1 obtained by distributing periodically Mn$^{3+}$ and
Mn$^{4+}$ ions.

\begin{figure}
  \includegraphics[clip,width=8cm]{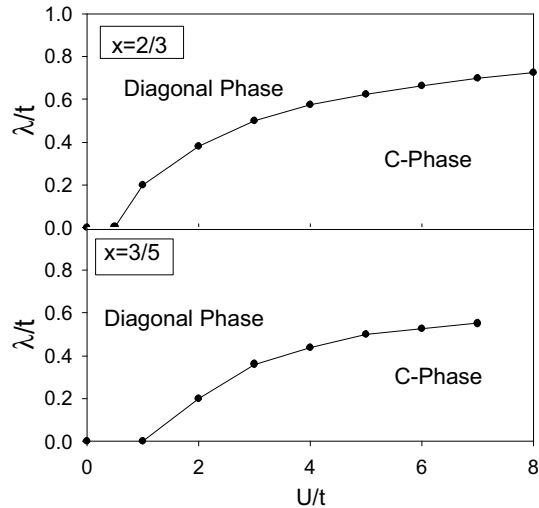}
  \caption{Phase diagram $(\lambda - U$),  for the three-dimensional DE
  two orbital model with cooperative Jahn Teller phonons, at hole concentrations
   $x$=2/3 and $x$=3/5.}
  \label{fig3}
\end{figure}
The phase diagram  above presented correspond to $U$=0. It is
known\cite{Shen} that at $x$=1/2 the $U$ interaction destabilizes
the CE phase against the C phase. For analyzing the estability of
the diagonal phase  we plot in Fig.3 the $\lambda$-$U$ phase
diagram for $x$=2/3 and $x$=3/5. In the C phase only an orbital
$3y^2 -r^2$ is occupied and the energy of this phase is
independent of $U$. Therefore the on site Coulomb interaction
favors the C phase against the diagonal phase and it is necessary
a finite electron phonon coupling to stabilize the diagonal phase.
However the require values of $\lambda$  are moderate and the
diagonal phase has lower energy than the C phase for realistic
estimation of this parameter\cite{Aliaga}.

\begin{figure}
  \includegraphics[clip,width=8cm]{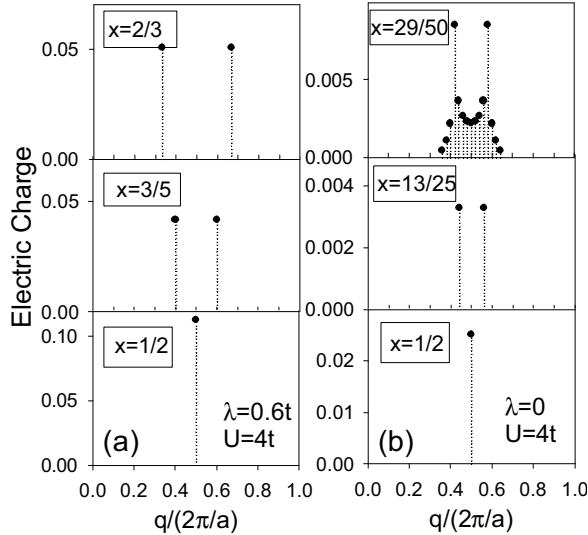}
  \caption{Diagonal, $(q,q)$ Fourier component of the electric charge in the diagonal phase.
  (a) corresponds to  $\lambda$=0.6$t$ and $U$=4$t$ and (b) to
  $\lambda$=0 and $U$=4t.}
\label{fig4}
\end{figure}
The results shown until here indicate the existence of a diagonal
phase at $x$=2/3 and $x$=3/5. As commented above the Hubbard term
and the electron phonon interaction favor the existence of a
predominant Fourier component, $\frac{m}{n}(\frac{2\pi
}{a},\frac{2\pi}{a})$ of the charge density. In Fig.4(a) we plot
the Fourier transform of the electrical charge for $x$=1/2,
$x$=2/3 and $x$=3/5 and $\lambda$=0.6$t$ and $U$=4$t$. Only the
Fourier components $\pm (\frac{2
\pi}{a},\frac{2\pi}{a})\frac{m}{n}$ have values considerably
different from zero.

There is experimental evidence\cite{Loudon2} that the period of
the charge modulation varies continuously  with the hole density.
The study of a system with a hole concentration close to half
filling, as $x$=0.52 or $x$=0.58,  requires the use of very large
unit cell and it is computationally unaccessible. However for zero
electron phonon coupling the properties of the diagonal phase can
be obtained by solving one dimensional system with $2m$ Mn ions
per unit cell. In this form we study systems with $x$ close to
1/2. In Fig.4(b) we plot the Fourier transform of the electron
charge for $U$=4t, $\lambda$=0 and hole densities $x$=1/2,
$x$=13/25 and $x$=29/50. It is noticeable that even in absence of
electron phonon coupling the charge density is diagonally periodic
with a prevailing Fourier component $\frac{m}{n}(\frac{2\pi
}{a},\frac{2\pi}{a})$. We expect that the inclusion of electron
phonon coupling will make more predominant this Fourier component.
These results indicate that the diagonal phase describes an state
where the electric charge is ordered diagonally and the
periodicity of the charge modulation changes continuously as
$a/(1-x)$.

\textit{Summarizing}, we propose a diagonal phase for manganites
at doping $x$=$(1-\frac{n}{m})$. In this phase the charge is
modulated diagonally with a prevalent Fourier component
$\frac{m}{n}(\frac{2\pi }{a},\frac{2\pi}{a})$. Magnetically this
phase consists of zig-zag chains with horizontal and vertical
steps formed by $m$+1 Mn ions. We obtain that this phase has lower
energy than the previous suggested phases and is the ground state
for a wide range of the relevant physical parameters. In agreement
with recent experimental results this phase describes a state
where the periodicity of the charge changes continuously with the
hole doping.

{\it Acknowledgments} I am grateful to F.Guinea, Mar Garcia,
M.P.L\'opez Sancho and N.Mathur for helpful discussions. Also I
thank H.Aliaga for calling my attention on the island phases.
Financial support is acknowledged from Grants No
MAT2002-04429-C03-01 (MCyT, Spain) and Fundaci\'on Ram\'on Areces.

%%%%%%%%%%%%%%%%%%%%%%%%%%%%%%%%%%%%%%%%%%%%%%%

%\bibliography{mia}

\begin{thebibliography}{27}
\bibitem[{not()}]{nota3}
\bibinfo{note}{The relation between topology and band structure in manganites
  has been discussed in reference \cite{Hotta3}}.
\end{thebibliography}

\end{document}